\documentclass[showpacs,preprint,aps]{revtex4}
\usepackage{graphicx}
\usepackage{dcolumn}
\usepackage{bm}
\begin{document}
\setcounter{page}{1}
\title
{Comment on ``Hadronic $^3$He$\eta$ production near threshold"}
\author
{K. P. Khemchandani$^1$, N. G. Kelkar$^2$ and B. K. Jain$^3$}
\affiliation{$^1$ Departamento de Fisica Teorica and IFIC, CSIC,
 Aptd. 22085, 46071 Valencia, Spain\\
$^2$ Departamento de Fisica, Universidad de los Andes,
Cra.1E No.18A-10, Bogota, Colombia\\
$^3$ Department of Physics, University of Mumbai,
Mumbai, India}
\begin{abstract}
Measurements of the differential and total cross sections for the 
$p d \rightarrow \,^3$He $\eta$ reaction at five energies were recently
reported [Phys. Rev. C {\bf 75}, 014004 (2007)] and comparisons with 
theoretical models were made. 
We point out that these comparisons involved 
a model based on ad hoc assumptions and hence the conclusions regarding 
the reaction mechanism as well as the role of
the higher partial waves drawn in the above work are misleading. 
Revised conclusions based on better model calculations are presented. 
\end{abstract}
\maketitle

The $p d \rightarrow \,^3$He $\eta$ reaction has been studied earlier 
near threshold \cite{mayer} as well as at a few higher energies 
\cite{bilger,cosy1}. The strong role of the $\eta$ nucleus final state 
interaction (FSI) near threshold was established first in \cite{wilkin} 
where the FSI was incorporated through an enhancement factor. Later on,
using few body equations within the finite rank approximation to
describe the $\eta$ - $^3$He elastic scattering, a good agreement with 
the threshold data was found in \cite{we31} using a three body reaction 
mechanism.
However, this two-step model, where 
the $\eta $ meson is produced via the $ pp\rightarrow \pi d$ and 
$\pi N\rightarrow \eta N$ reactions could not reproduce the 
forward peaked angular distributions at high energies \cite{we32}. 
In \cite{we32}, it was also shown that the claim in \cite{stenmark}, 
that the two-step model is successful in reproducing the angular distributions 
as well as the total cross sections at high energies was 
based on ad hoc assumptions related to 
the intermediate off-shell pion. 
Besides, the author in \cite{stenmark} included the FSI in an approximate 
way. 
The purpose of this note is to point out the 
erroneous conclusions reached in [Phys Rev. C {\bf 75}, 014004 (2007)] 
(henceforth referred to as \cite{adam}) based on comparisons with the 
two step model of \cite{stenmark} as well as present and discuss the 
results on the
role of the higher partial waves in the $\pi N \rightarrow \eta N$ scattering, 
which is an input for these calculations. 
Finally, we also compare our theoretical $\eta \,^3$He scattering length 
with the one deduced 
by the authors of \cite{adam}. 

In Fig. 1, we compare the data on the total cross sections 
for the $p d \rightarrow \,^3$He $\eta$ reaction with our 
results obtained using the two-step model for the production mechanism 
\cite{we31,we32}. 
This calculation incorporates an integral over all 
momenta of the intermediate off-shell particles in the two-step model, 
and includes the $\eta\,^3$He FSI through a solution of the few body 
equations in the FRA.
The results are shown for the FSI calculated with two 
different parameter sets for the $\eta N \rightarrow \eta N$ scattering, 
corresponding to scattering lengths (0.75,0.27)fm and (0.88,0.41)fm.
This model reproduces the data 
of Mayer et al. \cite{mayer} as well as that of the COSY-ANKE experiment 
\cite{anke} at very low energies, close to threshold. 
The model is however unable to reproduce the data away from threshold 
in Fig. 1.

In Fig. 2, we show the predictions of this model for the angular 
distributions calculated
with the $s$-wave alone and with higher partial waves for the
intermediate  $\pi N \rightarrow \eta N$ process. 
The theoretical angular distribution at threshold (T$_p$ = 891 MeV) 
is isotropic (see Fig. 4 in \cite{we31}). However, the present two step
model soon leads to a backward peaked cross section 
already at an excess energy of $Q = 5$ MeV.
The forward peaked distributions at high energies cannot be 
reproduced even after the inclusion of the higher partial waves. 
The shifting of the peak to the forward hemisphere with energy, as shown in
\cite{we32}, can arise if we restrict the propagation of the 
intermediate pion to small angles (less than 10 deg). Such a constraint is, 
however, ad hoc and unjustified. Besides, the magnitude of the cross
section gets highly underestimated in such a situation
as can be seen in Fig. 4 of \cite{we32}.  

\begin{figure}[ht]
\includegraphics[width=15cm,height=14cm]{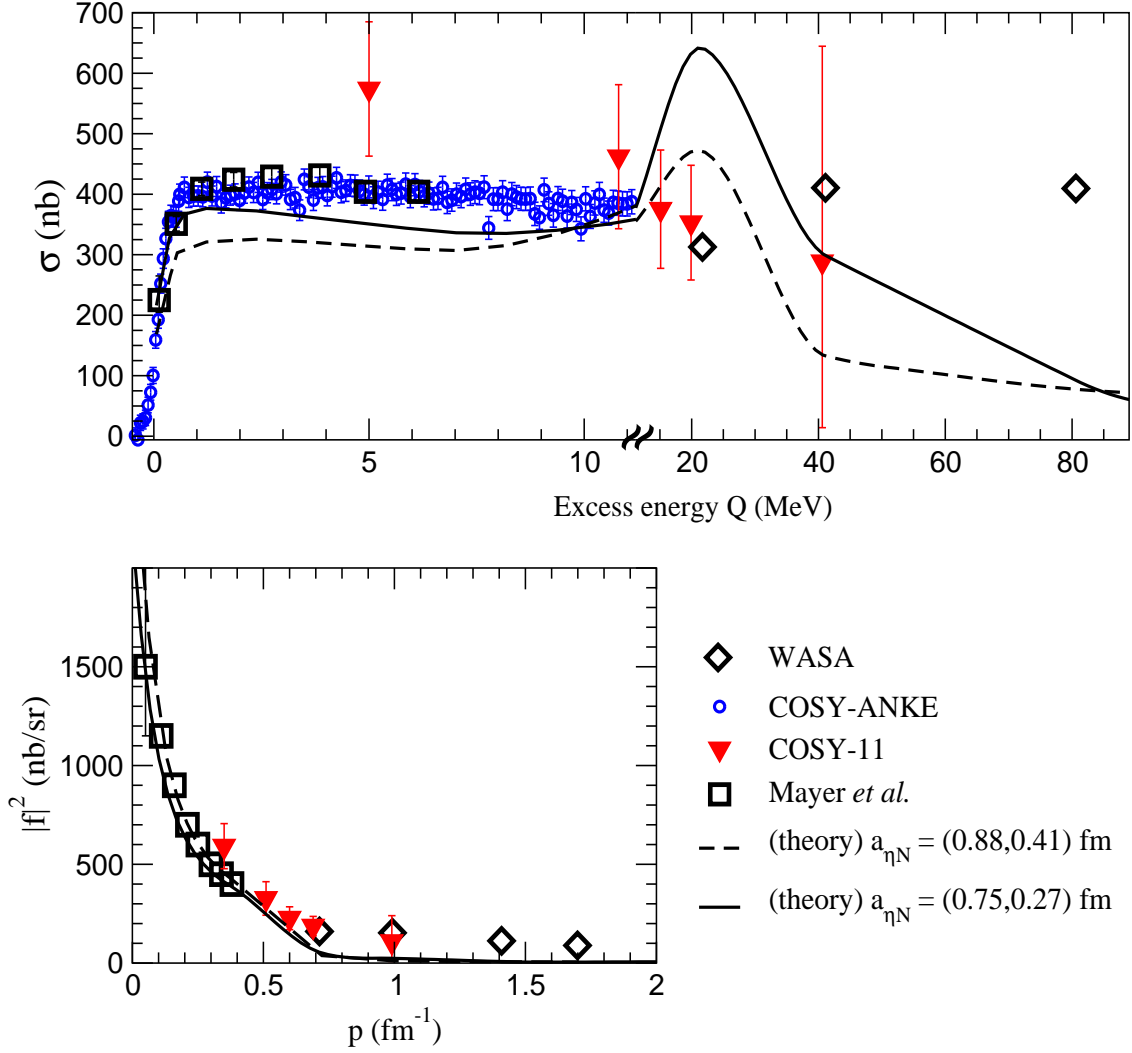}
\caption{\label{fig:eps1}
Comparison of the data on the $p d \rightarrow \,^3$He $\eta$ reaction 
with a two step model calculation including the 
$\eta \,^3$He final state interaction in the $s$-wave and $s$, $p$ and $d$ 
partial waves in the reaction mechanism. The scale in the upper plot has been
broken at 11.2 MeV for clarity.}
\end{figure}
\begin{figure}[h]
\includegraphics [width=11cm,height=10cm]{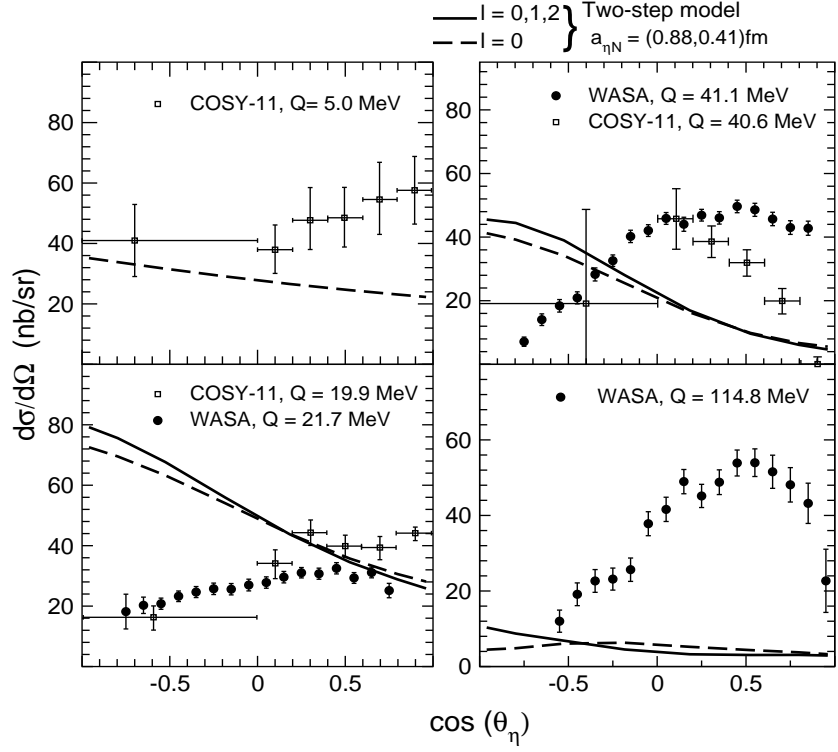}
\caption{\label{fig:eps2}
Role of the higher partial waves in the angular distributions for the 
$p d \rightarrow \,^3$He $\eta$ reaction (as explained in the text)
at various energies calculated within the two step model with FSI included.}
\end{figure}

Finally, the $\eta \,^3$He scattering lengths, 
calculated from the $\eta \,^3$He t-matrix in \cite{we31}, are  
$a_{\eta \,3He} = (1.99, 5.99)$ fm and $(2.14,5.71)$ fm for 
$a_{\eta \,N} = (0.75, 0.27)$ fm and $(0.88,0.41)$ fm, respectively, 
in comparison with the one reported in \cite{adam},viz., 
$a_{\eta \,3He} = (4.2\pm 0.5, 0.4\pm1.9)$ fm obtained from a fit to data. 
Our scattering lengths were 
obtained from the zero energy values of the few body 
$\eta \,^3$He $t$-matrix (with $s$-waves only) which
was also used in calculating the FSI of the $p d \rightarrow \,^3$He $\eta$ 
reaction. Though recent values of $|a_{\eta \,3He}|=4.3\pm0.5$ fm 
\cite{smyrski} and $a_{\eta \,3He} = (10.7\pm0.8, 1.5\pm2.6)$fm \cite{anke} 
are larger than our theoretical 
prediction, in agreement with the findings in \cite{anke}, 
evidence for a quasibound state very close to threshold 
was found in another of our works \cite{we33}
using the above $\eta ^3$He $t$-matrix corresponding to
an $\eta \,^3$He scattering length of $(2.14, 5.71)$fm.

To summarize, though the existing two-step model calculations of 
\cite{we31,we32},  
do reproduce the close to threshold,  
$p d \rightarrow \,^3$He $\eta$ 
total cross section data 
of Mayer \cite{mayer} and the COSY-ANKE experiment 
\cite{anke}, they are unable to reproduce the forward peaked angular
distributions at high energies. Hence any conclusions in \cite{adam} about 
the success of the two-step model, based on a 
comparison with the theoretical work in \cite{stenmark} should
be taken with caution.

\end{document}